\documentclass{article}
\usepackage{dcase2019,amssymb,amsmath,graphicx,url,times}
\usepackage{color}
\usepackage[utf8]{inputenc}
\usepackage{mathrsfs}
\usepackage[final]{microtype}
\usepackage{booktabs}
\usepackage{array}
\newcolumntype{P}[1]{>{\centering\arraybackslash}p{#1}}
\usepackage{enumitem}
\usepackage{soul}

\usepackage{cite}
\graphicspath{{./figs/}}

\title{Polyphonic Sound Event Detection and\\Localization using a Two-Stage Strategy}

\name{Yin Cao,$^{1*}$ Qiuqiang Kong,$^{1*}$\thanks{* Equal contribution.} Turab Iqbal,$^{1}$ Fengyan An,$^{2}$ Wenwu Wang,$^{1}$ Mark D. Plumbley$^{1}$}
\address{$^{1}$Centre for Vision, Speech and Signal Processing (CVSSP), University of Surrey, UK \\
\{yin.cao, q.kong, t.iqbal, w.wang, m.plumbley\}@surrey.ac.uk \\
$^{2}$School of Mechanical \& Automotive Engineering, Qingdao University of Technology, China\\ anfy@qut.edu.cn\\
}

\begin{document}

\ninept
\maketitle

\begin{sloppy}

\begin{abstract}
Sound event detection (SED) and localization refer to recognizing sound events and estimating their spatial and temporal locations. Using neural networks has become the prevailing method for SED. In the area of sound localization, which is usually performed by estimating the direction of arrival (DOA), learning-based methods have recently been developed. In this paper, it is experimentally shown that the trained SED model is able to contribute to the direction of arrival estimation (DOAE). However, joint training of SED and DOAE degrades the performance of both. Based on these results, a two-stage polyphonic sound event detection and localization method is proposed. The method learns SED first, after which the learned feature layers are transferred for DOAE. It then uses the SED ground truth as a mask to train DOAE. The proposed method is evaluated on the DCASE 2019 Task 3 dataset, which contains different overlapping sound events in different environments. Experimental results show that the proposed method is able to improve the performance of both SED and DOAE, and also performs significantly better than the baseline method.


\end{abstract}


\section{Introduction}
\label{sec:introduction}

Sound event detection is a rapidly developing research area that aims to analyze and recognize a variety of sounds in urban and natural environments. Compared to sound tagging, event detection also involves estimating the time of occurrence of sounds. Automatic recognition of sound events would have a major impact in a number of applications \cite{virtanen2018computational}. For instance, sound indexing and sharing, bio-acoustic scene analysis for animal ecology, smart home automatic audio event recognition (baby cry detection, window break alarm), and sound analysis in smart cities (security surveillance). 

Recently, approaches based on neural networks have been shown to be especially effective for SED \cite{dcase2016}. Unlike audio tagging problems \cite{choi2016automatic, xu2017convolutional, eduardo2018freesound}, which only aim to detect whether the sound events are present in a sound clip, SED also involves predicting temporal information of events. Early neural network architectures utilized fully-connected layers to detect temporally-overlapping sound events \cite{cakir2015polyphonic}. More recently, due to their success in image recognition, convolutional neural networks (CNNs) have become the prevailing architecture in this area \cite{piczak2015environmental, zhang2015robust, salamon2017deep, kong2019cross}. Such methods use suitable time-frequency representations of audio, which are analogous to the image inputs in computer vision. Another popular type of neural network is the recurrent neural network (RNN), which has the ability to learn long temporal patterns present in the data, making it suitable for SED \cite{parascandolo2016recurrent}. Hybrids containing both CNN and RNN layers, known as convolutional recurrent neural networks (CRNNs), have also been proposed, which have led to state-of-the-art performance in SED \cite{cakir2017convolutional, xu2017convolutional}.

Sound source localization, which focuses on identifying the locations of sound sources, on the other hand, has been an active research topic for decades \cite{brandstein2013microphone}. It plays an important role in applications such as robotic listening,  speech enhancement, source separation, and acoustic visualization. Unlike the dominance of neural-network-based techniques in SED, DOAE is mainly studied using two methods: parametric-based methods and learning-based methods.


Parametric-based DOAE methods can be divided into three categories \cite{brandstein2013microphone}: time difference of arrival (TDOA) estimation, maximized steered response power (SRP) of a beamformer, and high-resolution spectral estimation. Generalized cross-correlation (GCC) methods are the most widely-used approaches for TDOA estimation \cite{knapp1976generalized, scheuing2008correlation}. Since the TDOA information is conveyed in the phase rather than the amplitude of the cross-spectrum, the GCC Phase Transform (GCC-PHAT) was proposed, which eliminates the effect of the amplitude while leaving only the phase \cite{knapp1976generalized}. The primary limitation of parametric-based GCC methods is the inability to accommodate multi-source scenarios. 


Learning-based DOAE methods have the advantages of good generalization under different levels of reverberation and noise. They are designed to enable the system to learn the connections between input features and the DOA. There has already been a series of research addressing DOAE using deep neural networks \cite{xiao2015learning, vesperini2016neural, ma2017exploiting, chakrabarty2017broadband, chakrabarty2017multi, he2018deep, ferguson2018sound, sun2018indoor, adavanne2018direction}. Results show that they are promising and comparable to parametric methods. However, these neural-network-based methods are mainly based on static sources. In addition to spectrum-based features, GCC-based features, which can effectively supply time difference information, have also been used as the input features \cite{xiao2015learning, vesperini2016neural, he2018deep, ferguson2018sound, sun2018indoor}. In order to further improve learning-based methods, more practical real-world sources need to be considered.

In real-world applications, a sound event is always transmitted in one or several directions. Given this fact, it is reasonable to combine sound event detection and localization with not only estimating their respective associated spatial location, but also identifying the type and temporal information of sound. Therefore, it is worthwhile to study them together and investigate the effects and potential connections between them. Recently, DCASE 2019 introduced Task 3, which is Sound Event Localization and Detection (SELD) for overlapping sound sources \cite{dcase2019}. A recently-developed system known as SELDnet was used as the baseline system. SELDnet uses magnitude and phase spectrograms as input features and trains the SED and DOAE objectives jointly \cite{adavanne2018sound}. However, phase spectrograms are hard for neural networks to learn from, and further relationships between SED and DOAE have not been revealed.

In this paper, joint training of SED and DOAE is implemented first with log mel spectrograms and GCC-PHAT as the input features. According to the experimental results, SED is able to contribute to the performance of DOAE, while joint training of SED and DOAE degrades the performance of both. To solve this problem, a new two-stage method for polyphonic sound event detection and localization is proposed. This method deals with sound event detection and localization in two stages: the SED stage and the DOAE stage, corresponding to the SED branch and the DOAE branch in the model, respectively. During training, the SED branch is trained first only for SED, after which the learned feature layers are transferred to the DOAE branch. The DOAE branch fine-tunes the transferred feature layers and uses the SED ground truth as a mask to learn only DOAE. During inference, the SED branch estimates the SED predictions first, which are used as the mask for the DOAE branch to infer predictions. The experimental results show that by using the proposed method, DOAE can benefit from the SED predictions; both SED and DOAE can be improved at the same time. The proposed method performs significantly better than the baseline method.


The rest of the paper is organized as follows. In Section \ref{sec:proposed_method}, the proposed learning method is described in detail. Section \ref{sec:experiment} introduces the dataset used, other methods for comparison, and experimental results. Finally, conclusions are summarized in Section \ref{sec:conclusion}.

\begin{figure*}[ht]
  \centering
  \centerline{\includegraphics[width=\textwidth]{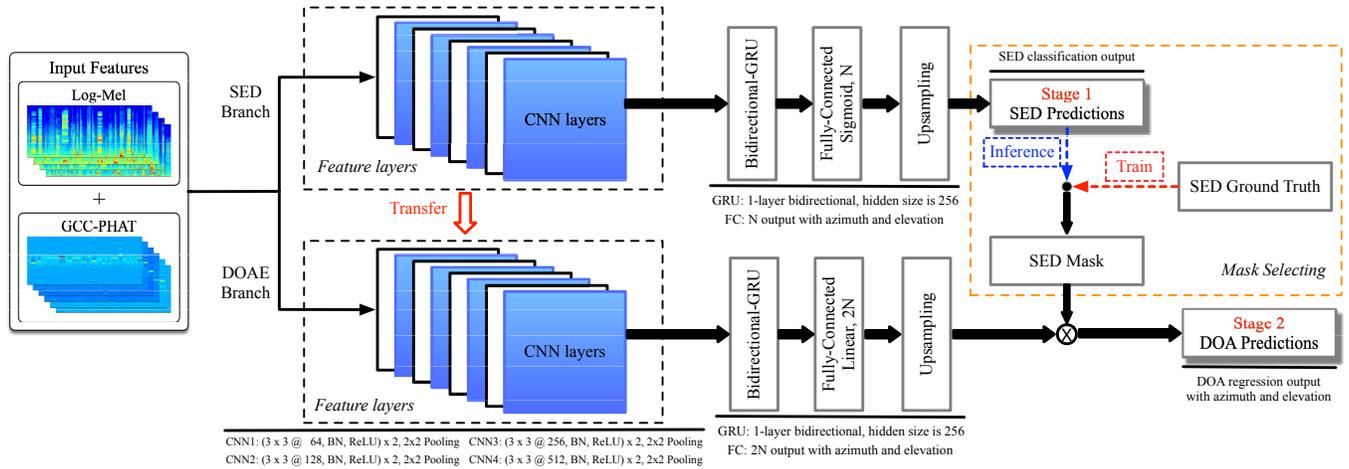}}
  \caption{The diagram of the proposed two-stage sound event detection and localization network. SED ground truth is used as the mask to train DOAE branch. SED predictions are used as the mask to infer DOAE.}
  \label{fig:network_architecture}
\end{figure*}

\section{Learning method}
\label{sec:proposed_method}

Joint training of SED and DOAE was first proposed in \cite{adavanne2018sound}. Their system is also used as the baseline system for DCASE 2019 Task 3. In this baseline, temporal consecutive magnitude and phase spectrograms are extracted as the input features from the time-domain audio waveform, which are then fed into a CRNN. Its loss is a weighted combination of the SED loss and the DOAE loss. Therefore, it can be imagined that this baseline system has an intrinsic trade-off between SED and DOAE according to the loss weight selected. In this paper, a two-stage polyphonic sound event detection and localization network is proposed to exploit their mutual strength.



\subsection{Features}
\label{ssec:input_features}

Selecting which features to use is an important factor for audio-related neural network applications. In this paper, the input signal format is of two types: First-Order of Ambisonics (FOA) or tetrahedral microphone array \cite{dcase2019}. Log mel spectrograms and GCC-PHAT, which contains phase difference information between all microphone pairs, are chosen as the input features. Ambisonics and GCC-PHAT are explained in this section.


\subsubsection{Ambisonics}
\label{sssec:ambisonics}

Ambisonics was developed as a spatial sound encoding approach several decades ago \cite{gerzon1985ambisonics}. It is based on the spherical harmonic (SH) decomposition of the sound field. Ambisonics encoding for plane-wave sound fields can be expressed as
\begin{equation}
\mathbf{b}(t)=\sum_{n=0}^{N}\mathbf{y}_{n} s_{n}(t),
\end{equation}
where $s_{n}(t)$ is the $n$-th plane-wave source signal, $N$ is the total number of sources, and $\mathbf{y}_{n}$ is the vector of the spherical harmonic function values for direction $(\theta_{n},\phi_{n})$, and can be expressed as
\begin{equation}
\begin{aligned} 
\mathbf{y}_{n}=&\left[Y_{0}^{0}\left(\theta_{n}, \phi_{n}\right), Y_{1}^{-1}\left(\theta_{n}, \phi_{n}\right), Y_{1}^{0}\left(\theta_{n}, \phi_{n}\right),\right.\\ 
&\quad  Y_{1}^{1}\left(\theta_{n}, \phi_{n}\right), \ldots, Y_{L}^{-L}\left(\theta_{n}, \phi_{n}\right), \ldots, Y_{L}^{0}\left(\theta_{n}, \phi_{n}\right), \\
&\quad \ldots, Y_{L}^{L}\left(\theta_{n}, \phi_{n}\right) ]^{T},
\end{aligned}
\end{equation}
where $L$ indicates the order of Ambisonics. It can be seen that Ambisonics contains the information of the source DOA. In addition, a higher directional resolution relates to a higher order of Ambisonics. Order-$L$ of Ambisonics needs at least $(L+1)^2$ microphones to encode. In real applications, the sound field is recorded using a spherical microphone array and converted into Ambisonics.



\subsubsection{Generalized Cross-Correlation}
\label{sssec:gcc}

GCC is widely used in TDOA estimation by means of maximizing the cross-correlation function to obtain the lag time between two microphones. The cross-correlation function is usually calculated through the inverse-FFT of the cross power spectrum. GCC-PHAT is the phase-transformed version of GCC, which whitens the cross power spectrum to eliminate the influence of the amplitude, leaving only the phase. GCC-PHAT can be expressed as 
\begin{equation}
GCC_{i j}(t, \tau)=\mathcal{F}^{-1}_{f \rightarrow \tau} \frac{X_{i}(f, t) X_{j}^{*}(f, t)}{\left|X_{i}(f, t) \| X_{j}(f, t)\right|},
\end{equation}
where $\mathcal{F}^{-1}_{f \rightarrow \tau}$ is the inverse-FFT from $f$ to $\tau$, $X_{i}(f, t)$ is the Short-Time Fourier Transform (STFT) of the $i$-th microphone signal, and $^*$ denotes the conjugate. TDOA, which is the lag time $\Delta \tau$ between two microphones, can then be estimated by maximizing $GCC$ with respect to $\tau$. Nevertheless, this estimation is usually not stable, especially in high reverberation and low SNR environments, and does not directly work for multiple sources. However, $GCC_{i j}(t, \tau)$ contains all of the time delay information and is generally short-time stationary. $GCC_{i j}(t, \tau)$ can also be considered as a GCC spectrogram with $\tau$ corresponding to the number of mel-band filters. That is, GCC-PHAT can be stacked with a log mel spectrogram as the input features. In order to determine the size of GCC-PHAT, the largest distance between two microphones $d_{\max}$ needs to be used. The maximum delayed samples corresponding to $\Delta \tau_{\max }$ can be estimated by $d_{\max}/c\cdot f_{s}$, where $c$ is the sound speed and $f_{s}$ is the sample rate. In this paper, log mel and GCC-PHAT are stacked as the input features, considering the possibility of the advance and the delay of GCC. The number of mel-bands, therefore, should be no smaller than the doubled number of delayed samples plus one \cite{knapp1976generalized}. 


\subsection{Network architecture}
\label{ssec:network}

The network is shown in Fig. \ref{fig:network_architecture}, and has two branches, the SED branch and the DOAE branch. During training, the extracted features, which have shape $C \times T \times F$, are first sent to the SED branch. $C$ indicates the number of feature maps, $T$ is the size of time bins, and $F$ is the number of mel-band filters or delayed samples of GCC-PHAT. The CNN layers, which are also named as feature layers in this paper, are constructed with 4 groups of 2D CNN layers (Convs) with $2 \times 2$ average-pooling after each of them.  Each Convs' group consists of two 2D Convs, with a receptive field of $3 \times 3$, a stride of $1 \times 1$, and a padding size of $1 \times 1$ \cite{kong2019cross}. The Convs' kernels are able to filter across all of the channels of the input features or the feature maps from the last layer, hence are able to learn inter-channel information. CNN layers are capable of learning local temporal and frequency information to better abstract the event-level information. Each single CNN layer is followed by a Batch Normalization layer \cite{ioffe2015batch} and a ReLU activation. After the CNN layers, the data has shape $C_{out} \times T/16 \times F/16$, where $C_{out}$ is the number of output feature maps of the last CNN layer. It is then sent to a global average-pooling layer to reduce the dimension of $F$. After this, the data is reshaped to have shape $T/16 \times C_{out}$ and is fed to a bidirectional GRU. The output size is maintained and is sent to a fully-connected layer with output size $T/16 \times N$, where $N$ is the number of event classes. The sigmoid activation function is used afterwards with an upsampling in the temporal dimension to ensure the output size is consistent with $T$. The SED predictions can then be obtained through an activation threshold. Binary cross-entropy is used for this multi-label classification task. 

The DOAE branch is then trained. The CNN layers are transferred from the SED branch and are fine-tuned. The output of the fully-connected layer for the DOAE branch is a vector of $N \times 2$ linear values, which are azimuth and elevation angles for $N$ events. They are then masked by the SED ground truth during training to determine if the corresponding angles are currently active. Finally, the mean absolute error is chosen as the DOAE regression loss.

During inference, the SED branch first computes the SED predictions, which are then used as the SED mask to obtain the DOAE. The reason for building this network architecture is to enhance the representation ability of a single network so that each branch is only responsible for one task, while the DOAE branch can still incorporate the benefits contributed from SED. 


\section{Experimental Study}
\label{sec:experiment}

The proposed two-stage polyphonic sound event detection and localization method is compared with other methods described in Section \ref{ssec:compared_methods}. They are evaluated on the DCASE 2019 Task 3 dataset \cite{dcase2019}. This task is for sound event detection and localization. The dataset provides two formats of data: 1) First-Order of Ambisonics; 2) tetrahedral microphone array. The development set consists of 400 one minute long recordings, divided into four cross-validation splits. There are 11 kinds of isolated sound events in total. The audio recordings are mixtures of isolated sound events and natural ambient noise. The sound events are convolved with impulse responses collected from five indoor locations, resulting in 324 unique combinations of azimuth-elevation angles. One challenging problem in this dataset is that the sound events in the audio recordings have a polyphony of up to two, which means sound events from different locations may overlap. The source code for this paper is released on GitHub\footnote{https://github.com/yinkalario/Two-Stage-Polyphonic-Sound-Event-Detection-and-Localization}.


\subsection{Evaluation metrics}
\label{ssec:eval_metrics}

Polyphonic sound event detection and localization are evaluated with individual metrics for SED and DOAE. For SED, segment-based error rate (ER) and F-score\cite{mesaros2016metrics} are calculated in one-second lengths. A lower ER or a higher F-score indicates better performance. In addition, mean average precision (mAP), which is the area under the precision and recall curve, is used to evaluate the frame-level tagging performance. The mAP is used here because it does not depend on the threshold selection, hence is able to better objectively evaluate the performance. A higher mAP indicates better performance. For DOAE, DOA error and frame recall are used \cite{adavanne2018direction}. A lower DOA error or a higher frame recall indicates better performance.







\subsection{Methods for comparison}
\label{ssec:compared_methods}

In order to show the effectiveness of the proposed method, several other methods are compared, including
\begin{itemize}[itemsep=0.1ex, leftmargin=0.3cm, label=\raisebox{0.25ex}{\tiny$\bullet$}]
    \item \textbf{Baseline}, which is the baseline method used in DCASE 2019 Task 3, uses magnitude and phase spectrograms as the input features. The features are then fed to a CRNN network. The loss of SED and DOAE are combined and jointly trained. 
    \item \textbf{SELDnet}, which has the same architecture with the baseline but using log mel and GCC-PHAT spectrograms as the input features.
    
    \item \textbf{DOA}, which uses log mel and GCC-PHAT spectrograms as the input features to only estimate DOA. It transfers the CNN layers from the SED network and utilizes SED ground truth as the mask.
    
    \item \textbf{DOA-NT}, is the same as DOA method except for not transferring CNN layers. Both DOA and DOA-NT only estimate DOAs.
\end{itemize}
All of the above-mentioned methods are evaluated on both CNNs and CRNNs. The CNN has the same architecture as the CRNN but without the recurrent layer. Furthermore, microphone array signals do not need extra encoding processes. It is more convenient to use in practice, whereas the encoding of FOA may contain extra spatial information. Therefore, it is worthwhile to evaluate these methods with both the microphone array and FOA data.

\begin{table}
  \caption{Performance for the development dataset.}
  \vspace{6pt}
  \label{tab:score_result}
  \centering
  \resizebox{\columnwidth}{!}{%
  \begin{tabular}{l l P{0.45cm}P{0.45cm}P{0.45cm}P{0.45cm}P{0.45cm}P{0.45cm}P{0.45cm}P{0.45cm}P{0.45cm}P{0.45cm}}
  \toprule
    \multicolumn{2}{l}{}    & \multicolumn{5}{c}{\textbf{\textsc{mic-array}}}  &\multicolumn{5}{c}{\textbf{\textsc{FOA}}}\\
    \cmidrule(lr){3-7} \cmidrule(lr){8-12} 
    Methods & Net   & ER    & F     & mAP   & DOA               & FR    & ER    & F     & mAP   & DOA           & FR    \\
  \midrule
    Baseline& CRNN  & 0.350 & 0.800 & $-$   & 30.8$^{\circ}$    & 0.840 & 0.340 & 0.799 & $-$   & 28.5$^{\circ}$& 0.854 \\
  \midrule
    SELDnet & CNN   & 0.277 & 0.844 & 0.718 & 11.0$^{\circ}$    & 0.827 & 0.281 & 0.843 & 0.718 & 10.9$^{\circ}$& 0.828 \\
            & CRNN  & 0.213 & 0.879 & 0.770 & 11.3$^{\circ}$    & 0.847 & 0.221 & 0.876 & 0.768 & 12.6$^{\circ}$& 0.844 \\
  \midrule
    DOA     & CNN   & $-$   & $-$   & $-$   & 13.3$^{\circ}$    & $-$   & $-$   & $-$   & $-$   & 13.1$^{\circ}$& $-$   \\
            & CRNN  & $-$   & $-$   & $-$   & 11.9$^{\circ}$    & $-$   & $-$   & $-$   & $-$   & 11.9$^{\circ}$& $-$   \\
  \midrule    
    DOA-NT  & CNN   & $-$   & $-$   & $-$   & 14.7$^{\circ}$    & $-$   & $-$   & $-$   & $-$   & 14.5$^{\circ}$& $-$   \\
            & CRNN  & $-$   & $-$   & $-$   & 14.0$^{\circ}$    & $-$   & $-$   & $-$   & $-$   & 14.3$^{\circ}$& $-$   \\
  \midrule    
    Two-Stage & CNN   & 0.251 & 0.862 & 0.749 & 10.9$^{\circ}$    & 0.832 & 0.248 & 0.864 & 0.756 & 10.8$^{\circ}$& 0.832 \\
            & CRNN  &\textbf{0.167}&\textbf{0.909}&\textbf{0.819 }& \textbf{9.85$^{\circ}$}& \textbf{0.863} &\textbf{0.181}&\textbf{0.898}&\textbf{0.800}& \textbf{9.84$^{\circ}$}& \textbf{0.857} \\
  \bottomrule
  \end{tabular}
  }
\end{table}


\subsection{Hyper-parameters}
\label{ssec:parameters}

To extract the input features, the sample rate of STFT is set to 32kHz. A 1024-point Hanning window with a hop size of 320 points is utilized. In the DCASE 2019 Task 3 dataset, the largest microphone distance is 4.82cm \cite{dcase2019}. According to Section \ref{sssec:gcc}, the number of mel-band filters and the delays of GCC-PHAT is set to be 64. For 4 channels of signals, up to 10 input channels of signals are sent to the network. The audio clips are segmented to have a fixed length of 2 seconds with a 1-second overlap for training. The learning rate is set to 0.001 for the first 30 epochs and is then decayed by 10\% every epoch. The final results are calculated after 50 epochs.


\begin{figure}[ht]
  \centering
  \centerline{\includegraphics[width=\columnwidth]{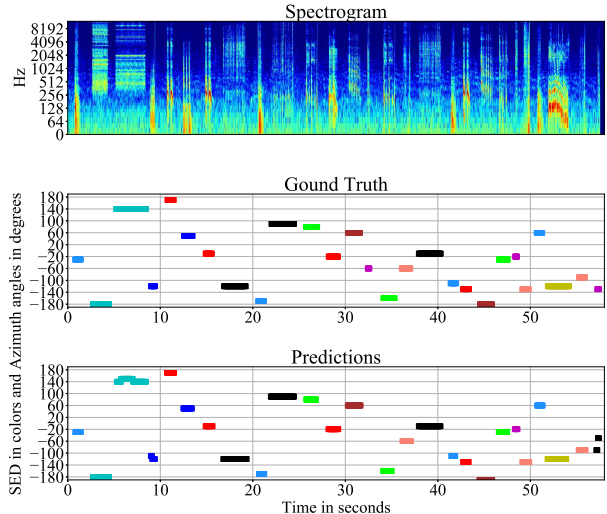}}
  \caption{SED and DOAE Azimuth results for proposed two-stage method. Different colors indicate different classes of events.}
  \label{fig:SED_DOA_resutls}
\end{figure}

\subsection{Results}
\label{ssec:results}

The experimental results are shown in Table \ref{tab:score_result}. SELDnet with log mel and GCC-PHAT spectrograms as the input features was implemented first to compare with the baseline method. It can be seen from both microphone array data and FOA data that with log mel and GCC-PHAT spectrograms as the input features, SELDnet outperforms the baseline system using magnitude and phase spectrograms. Log mel spectrograms are more effective input features than magnitude spectrograms, not only due to their better performance, but they are also more compact. GCC-PHAT spectrograms, which mainly contain the time difference information, show their advantages over phase spectrograms. The results of DOA and DOA-NT show that with trained CNN layers transferred, DOA error is consistently lower than not transferring, which indicates that SED information contributes to the DOAE performance; it can also be observed that the convergence speed is much faster with CNN layers transferred. Comparing SELDnet with DOA-NT, it also shows that the joint training is better than the training of DOAE without CNN layers transferred, which also proves SED contributes to DOAE. The proposed two-stage method is presented in the end. The metrics scores are the best among all the methods. Compared with SELDnet, it indicates that the joint training of SED and DOAE degrades the performance of both. This two-stage method minimizes each loss individually, hence the network representation ability is enhanced for each sub-task, while the contribution from SED to DOAE is still preserved by transferring CNN layers to the DOAE branch.

Comparing microphone array data and FOA data, the results do not show FOA is better, which means FOA does not necessarily contain more spatial information than microphone array signals with four channels. On the other hand, in most cases, CRNNs perform better than CNNs, which indicates that long temporal information may be useful for both SED and DOAE. A visualization of SED and DOAE using the proposed method for one clip is shown in Fig. \ref{fig:SED_DOA_resutls}. It can be seen that most of the SED and DOAE predictions are accurate in both temporal and spatial dimensions.


\section{Conclusion}
\label{sec:conclusion}

Treating sound event detection and localization as a combined task is reasonable. In this paper, it shows that SED information can be used to improve the performance of DOAE. However, joint training of SED and DOAE degrades the performance of both. A two-stage polyphonic sound event detection and localization method is proposed to solve this problem. The proposed method uses log mel and GCC-PHAT spectrograms as the input features and has two branches of SED and DOAE. The SED branch is trained first, after which the trained feature layers are transferred to the DOAE branch. The DOAE branch then uses the SED ground truth as a mask to train DOAE. Experimental results show that the proposed method is able to enhance the network representation ability for each branch, while still keeping the contributions from SED to DOAE. The proposed method is shown to significantly outperform the baseline method.

\section{ACKNOWLEDGMENT}
\label{sec:ack}

This research was supported by EPSRC grant EP/N014111/1 ``Making Sense of Sounds'' and was partially supported by the H2020 Project
entitled AudioCommons funded by the European Commission with Grand Agreement number 688382.


\bibliographystyle{IEEEtran}
\bibliography{refs19}

\end{sloppy}
\end{document}